\documentstyle[12pt]{article}
\def\abstracts#1{{
\centering{\begin{minipage}{12.2truecm}\footnotesize
\baselineskip=12pt\noindent
\centerline{\footnotesize ABSTRACT}\vspace*{0.3cm}
\parindent=0pt #1
\end{minipage}}\par}}

\renewenvironment{thebibliography}[1]
{\begin{list}{\arabic{enumi}.}
{\usecounter{enumi}\setlength{\parsep}{0pt}
\setlength{\leftmargin 1.25cm}{\rightmargin 0pt}
 \setlength{\itemsep}{0pt} \settowidth
{\labelwidth}{#1.}\sloppy}}{\end{list}}
\newcounter{itemlistc}
\newcounter{romanlistc}
\newcounter{alphlistc}
\newcounter{arabiclistc}

 1
 1
 1

\newcommand{\be}{\begin{eqnarray}}
\newcommand{\ee}{\end{eqnarray}}
\begin{document}
\begin{flushright}
{\normalsize UAHEP953}
\end{flushright}
\centerline{\normalsize\bf Top Quark Candidates and Light Gluinos$^{*}$}
\vspace*{0.6cm}
\centerline{\footnotesize L. Clavelli}
\baselineskip=13pt
\centerline{\footnotesize\it Dept. of Physics and Astronomy,
    University of Alabama}
\centerline{\footnotesize\it  Tuscaloosa AL 35487}
\centerline{\footnotesize  E-mail: lclavell@ua1vm.ua.edu}
\vspace*{0.9cm}
\baselineskip=12pt
\abstracts{ We discuss the case for a light gluino and the mutual
  impact of this possibility with the Fermilab top quark candidates.}
\vspace*{0.6cm}
\normalsize\baselineskip=15pt
\setcounter{footnote}{0}
\renewcommand{\thefootnote}{\alph{footnote}}

     In the last year or two, the Fermilab collider experiments, CDF
and D0, have observed several dozen isolated, high transverse momentum
leptons associated with significant jet activity and large missing
transverse momentum.  These events are highly improbable
(likelihood $~10^{-5}$) from the standpoint of previously known
particles and hence constitute clear evidence for either the top
quark or particles beyond the standard model.  In the same period
several anomalies at the $Z$ scale and below have lent support to the
idea that the gluino of supersymmetry (SUSY) lies in the low energy
region.  The strong mutual impact between this idea and the top quark
interpretation of the Fermilab events is the subject of this talk.
\par
It has long been known that there are one or more low energy
windows~\cite{UA1} allowed by experiment for the gluino mass.  In the
last few years, however, many authors have gone beyond this to point
out positive, though indirect, hints that this scenario may, in fact,
be realized in nature.  One can count at least seven such indications
of a light gluino.  These are:
\par
1.) {\bf Consistency of SUSY grand unified theory (GUT) with low
energy data on $\alpha_s(M_Z)$}.  The minimal SUSY GUT (MSSM)
predicts~\cite{{LP94},{Wright94},{CC95}} an $\alpha_s(M_Z)$ above
$0.117$ while most of the low energy data prefers a much smaller value.
The $e^+e^-$ jet measures, whose energy dependence indicates absence of
non-perturbative corrections, as well as a large body of quarkonium
data interpreted in the standard model, predict
$\alpha_s(M_Z) = .098\pm.003$.~\cite{CCY} The QCD sum rules which
provide a systematic way to control
non-perturbative corrections predict~\cite{Voloshin}
$\alpha_s(M_Z)=.109\pm.001$.
The truth may be somewhere between these two predictions but, in any
case, low energy analyses systematically prefer an anomalously low
value of $\alpha_s(M_Z)$ compared to GUT predictions. The problem can
be eliminated if the gluino lies at the $Z$ scale
or below~\cite{{Clavelli},{RS95}} while the other SUSY particles
apart from the photino are much heavier.  Unfortunately, this
solution\vadjust{
\vskip 2 mm
{\footnotesize $^{*}$ Talk presented at the Workshop on the Physics of
the Top Quark, Iowa State University, May 1995\hfill}
\pagebreak
}
violates the supergravity inspired mass relations among the
SUSY particles which we take for present purposes to be part of the
MSSM assumption.  Namely, a universal gaugino mass predicts in the
light gluino case a simultaneously light ($\simeq M_W$) gaugino
spectrum.  This is important in the discussion to follow of the
$b$ excess in $Z$ decay but causes the MSSM prediction of
$\alpha_s(M_Z)$ to again become high even in the presence of the light
gluino.~\cite{CCMNW} It now seems that if one wants to preserve the
supergravity mass universality conditions one cannot unambiguously
reach low values of $\alpha_s(M_Z)$ in the MSSM.  This has lent
support to the SUSY Missing Doublet Model (MDM) where low values of
$\alpha_s(M_Z)$ are naturally predicted~\cite{{CC95},{BMP}}.
Of course one can appeal to model-dependent non-renormalizable or
gravitational effects~\cite{RUA} but it seems then that it would be
as easy to destroy the good SUSY prediction of $\sin^2(\theta)$ as
to achieve a lower $\alpha_s(M_Z)$.
\par
     2.) {\bf Consistency of $\alpha_s(M_c)$ with
$\alpha_s(M_b)$.}~\cite{CCY} In the non-relativistic color singlet
model the values of the strong coupling
at these two scales are inconsistent by many standard deviations.  The
situation is greatly alleviated if the gluino lies in the low mass
window below 0.7 GeV.~\cite{CCY}  Alternatively, the problem might be
resolved by relativistic corrections but introducing an ad-hoc
parameter to represent these corrections~\cite{Kobel} implies a
$70\%$ reduction in the $J/\Psi$ decay rate due to this parameter,
leads to several paradoxes, and only succeeds in raising
$\alpha_s(M_Z)$ to about $0.113$ even with this $70\%$
model-dependent reduction.
\par
     3.) {\bf Consistency of $\alpha_s(M_Z)$ from direct measurements
at the $Z$ scale with that inferred from low energy measurements}.
The LEP measurements of the Z width interpreted in the standard model
lead to an $\alpha_s(M_Z) = 0.125\pm.005$.  This is inconsistent with
low energy measurements which, interpreted in the Standard Model as
discussed above, imply a value near or below $0.11$.  A gluino
in the $\Upsilon$ region or below causes $\alpha_s$  to run more
slowly and alleviates the problem.~\cite{{CCY},{JK}}  However, in the
non-relativistic quarkonium analysis,~\cite{CCY} even with the light
gluino effect one only reaches an $\alpha_s(M_Z)$ of $0.113$ implying
there is some extra contribution to the $Z$ width to make the apparent
$\alpha_s(M_Z)$ as great as $0.125$.  It has been suggested that this
extra contribution is associated with the $b$ excess in $Z$
decay~\cite{Rb} which is also natural in the light gluino case.
The problem has been known since $1992$ when an attempt was made to
explain the results in terms of $Z$ decays into
${\overline q}{\tilde q}{\tilde G}$ with a light gluino and a squark
mass between $M_Z/2$ and $M_Z$.~\cite{CCFHPY} This explanation is
still viable but it would not explain the $b$ excess in $Z$ decay
unless the $b$ squark is significantly lighter than the other squarks.
\par
 4.) {\bf Self-consistency of $Z$ data (width vs jet
 measures).}~\cite{ENR} The light gluino affects the jet measures
differently than the total width and improves consistency.
\par
     5.) {\bf Deep inelastic vs $Z$ scale measurement of
$\alpha_s(M_Z)$.}~\cite{BB}  In this analysis, however, the conclusion
that a light gluino is favored depends on the apparent high value of
$\alpha_s(M_Z)$ being correct.
\par
     6.) {\bf $b$ excess in $Z$ decay.}  The observed $b$ excess in $Z$
decay can be fit in SUSY models if the chargino and/or charged Higgs and
the stop quark are light (in the $50 $ GeV mass
region).~\cite{{WKK},{GJS}}  This cannot be achieved in the
supergravity inspired MSSM if the gluino is
heavy.  However in the model with a light gluino, the charginos and
neutralinos are necessarily in this region and the stop is also
naturally light if the universal scalar mass is below several
hundred GeV.~\cite{Rb}  The $b$ excess can be quantitatively
correlated with the apparently large value of $\alpha_s(M_Z)$
at LEP but even if the $b$ excess disappears with
more data, the large apparent $\alpha_s(M_Z)$ at LEP is an
independent indication that there are non-standard contributions to
the $Z$ width strongly suggesting low mass SUSY particles.   \par
  7.) {\bf Anomaly in the $B$ semi-leptonic branching ratio.}
There is a $2\sim 3~ \sigma$ discrepancy in the $B$ semi-leptonic
branching ratio which hints of non-standard model contributions to
$B$ non-leptonic decay.  It has been suggested~\cite{Kagan} that
the $B \rightarrow sG$ decay can be enhanced without unduly enhancing
$B \rightarrow s \gamma$ if the gluino is light.
\par
     Returning to indication $6$ above, we note that the squared
squark masses in the MSSM, although dominantly determined by the
squared universal scalar mass, $m_0^2$, receive a positive
contribution proportional to the squared universal
gaugino mass, $m_{1/2}^2$. Unless this parameter is negligible it is
impossible to achieve a sufficiently light stop quark mass to explain
the $b$ excess.  Although abandoning the supergravity inspired mass
relations has been suggested, these relations are worth defending
and still viable in the light gluino case where $m_{1/2} =0$
(minimal gauge-kinetic coupling).  To illustrate this viability we
put $m_{1/2}=0$ and run a Monte-Carlo choosing random values of the
five parameters $m_0$, $\mu$, $m_t$, $\tilde A_t$ , and $\tan \beta$.
Here $m_t$ is the top mass, $\mu$ is the Higgs mixing
parameter, $\tilde A_t$ is the dimensionless top trilinear coupling,
and $\tan \beta$ is the ratio of the Higgs vacuum expectation values.
We reject solutions with chargino or neutralino masses too low for
the LEP data.  We impose the $\rho$ parameter result
$\rho=1.009\pm.002$ and if $m_t <158 $ GeV we require that $m_{\chi}
+m_b <m_{\tilde t} < m_t$.  This latter constraint is used since it
allows prominent top decays into light gluinos~\cite{CC95} thus
evading the Fermilab high $P_T$ lepton trigger.  In this case, of
course, one has to find an alternate (SUSY) explanation
for the top quark candidates.  We find experimental consistency with
the MSSM and light gluinos if $45$ GeV $< m_{\chi^\pm}  < 52$ GeV,
$1.5 < \tan \beta < 1.74$, $52$ GeV $< \vert\mu\vert < 78$ GeV, and
$124$ GeV $< m_t  < 182$ GeV.  In fig.~1 we show the Monte-Carlo
solutions projected into the $m_t,m_{\tilde t}$ plane where
${\tilde t}$ is the lightest stop mass.
Solutions with a sufficient $b$ excess in $Z$ decay due to
stop-chargino loops lie below and to the left of the dashed line.  In
the heavy gluino case solutions below $M_t=158$ GeV are excluded by
the Fermilab results and the ${\tilde t}$ mass is raised due to the
contribution from $m_{1/2}$ so no solutions remain below the dashed
line.
\par
The $\mu$ parameter is severely constrained by the gaugino masses.
If radiative breaking of the electroweak symmetry holds, its square
must be equal to
\be
 \mu_{rad}^2 = - M_{Z}^2/2 - m_0^2{\bigl [}1 +
 {{(m_t/205 {\rm GeV})^2} \over{2 \cos(2 \beta)}}{\bigl (}3
 + {\tilde A_t}^2
 {\bigl \{}1-{{(m_t/205 {\rm GeV})^2} \over {\sin^2 \beta}}
 {\bigr \}}{\bigr )}{\bigr ]}.
						     \label{eq:mu}
\ee
Since this is a perturbation theory result we might require only that
\be
-0.15<{{\vert \mu_{rad} \vert - \vert \mu \vert} \over
  {\vert \mu_{rad} \vert + \vert \mu \vert} } < 0.15 .  \label{eq:mu2}
\ee
Solutions that satisfy this inequality
are indicated by $x$'s in fig.~1.  If the two loop contributions
to the light Higgs mass are negligible, this radiative breaking
constraint may disfavor the light gluino scenario by
predicting too low a Higgs mass (in the $50 $ GeV region)~\cite{LNW}.
However, a new understanding of dominant light Higgs decay modes into
gluinos~\cite{DD} might affect those conclusions.  The top quark mass
upper limits in fig. 1 are driven by the $\rho$ parameter constraint.
Top masses above $182$ are inconsistent with the measured value of
the $\rho$ parameter in the SUSY case with or without light gluinos.
If the top quark is above $160$
GeV and decays into $Wb$ with a branching ratio near unity then either
the light gluino hypothesis or that of radiative breaking with a
universal GUT scale Higgs mass is ruled out.  If the gluino is heavy
the\vadjust{
\vskip 3.5 in
\centerline {\footnotesize Fig.1. Allowed Values of $M_t$ and
$M_{\tilde t}$ in the Light Gluino Case}
\vskip 0.3 in
}
explanation of the $b$ excess through stop-chargino or
stop-charged-Higgs is wrong unless the supergravity related mass
relations are broken.  The radiative breaking prediction as
currently derived is dependent on a universal scalar mass.
However, it might be
considered less problematic to disturb the universality of Higgs
masses at the GUT scale than that of the gaugino or sfermion masses
due to the fact that some of these latter lie in the same $SU(5)$
multiplets.~\cite{CCMNW} \par
     It can also be seen from the $x$'s in fig.~1 that the
radiative breaking
constraint prefers solutions that lie below the dashed
line giving therefore a $b$ excess in $Z$ decay.  However, in this
case we must seek an alternate (SUSY) explanation for the FNAL
events and explain how a lower mass top could have been missed.
Such an explanation might run as follows.
\par
     1)  The background from $W +$ jets could be substantially
increased in the light gluino scenario due to the extra light
color octet.  The $W + 4$ jets sample would include
$WGG{\tilde G}{\tilde G}$ and $W{\tilde G}{\tilde G}{\tilde G}
{\tilde G}$ closely related to the $WGGGG$ final state
of the standard model.  To what extent the kinematics of these
final states would mimic the CDF and D0 top quark candidates is
at present unknown.  The CDF collaboration proposes to limit
any `non-standard' $W+$ multi-jet background by studying the
measured $Z+$ multi-jet events which are not expected to have
a contribution from top decay.~\cite{CDF}  In its $1994$ paper CDF
found, however, a $Z+\ge 3$ jet signal with $b$ tagging
$3.1\pm0.3$ times the expected background.  Although this is reduced
in its $1995$ paper, it is still consistent with twice the expected
background. Thus, the published CDF results on $Z +$ multi-jets
do not exclude a $W +$ multi-jet background at a level of twice the
standard model background. D0~\cite{D095} also notes an anomalously
large $W +$ jet production manifested by a fit value of $\alpha_s$
significantly greater than expected.  If there are larger than
expected $W +$ jet backgrounds, the likelihood that the top
candidates are statistical fluctuations from non-top backgrounds is
very much greater than the quoted $10^{-5}$ but there still may be
a need for some leptonic decays of new particles perhaps as follows.
\par
  2)  In the light gluino case, it is possible to produce heavy
squarks in association with a light gluino rather than with another
heavy sparticle. This enhances the cross section for a number of
processes which could enter the CDF and D0 samples.  A prime example
is the process $p \overline{p} \rightarrow \overline{b}{\tilde b}
{\tilde G}$.
This could be important even if the ${\tilde b}$ is above $200 $ GeV
since it is produced singly. The ${\tilde b}$ could have a large
branching ratio into ${\tilde t} W$ with the ${\tilde t}$ subsequently
decaying into $b {\overline e}\nu {\tilde \gamma}$ through a virtual
(or real) chargino.  Another process which could affect the top mass
determination is the production of ${\tilde t}{\tilde {\overline t}}$
or ${\overline t}{\tilde t}{\tilde G}$.  The first of these has been
already considered~\cite{ACO} recently.  The presence of the extra
(gluino) jet in the latter process raises the question whether the CDF
analysis has a bias toward large top quark mass due to their
using only the four jets of highest energy.\par
  A lower mass top could have been missed if its non-$W$ decays such as
${\tilde t}{\tilde G}$ have a significant branching ratio.  In fact one
would expect
\be
   {{\Gamma(t \rightarrow {\tilde t}{\tilde G})}
   \over{\Gamma(t \rightarrow
     b W)}} ={{2 \alpha_s} \over {3 \alpha}} \sin^2(\theta)
     {{\bigl (}{{M_W} \over {M_t}}{\bigr )}}^2
      {{{(1 - M_{\tilde t}^2/M_t^2)^2} \over
      {(1 - M_W^2/M_t^2)^2}}} (1+2 M_W^2/M_t^2)^{-1} . \label{eq:GT}
\ee
Thus these modes could be comparable in the light gluino, light stop
case. In the events where both tops decay to $bW$, there is a
two-fold ambiguity in top mass due to the choice of which $b$ jet
to associate with which $W$.  There should be a low mass solution
and a high mass solution both of which would lead to approximately
equal masses of the two tops if the tops are produced with
significant transverse momentum as is likely.
\par
In conclusion it seems that the published top quark analyses, using
Monte Carlos which are heavily dependent on the standard model, might
not adequately treat the possibility that a low mass top could be
hiding under the $W +$ multi-jet background in this light gluino
scenario.
\par
This work was supported in part by the DOE under grant
DE-FG05-84ER40141.

\end{document}